\documentclass[journal]{IEEEtran}
\usepackage{cite}     
\usepackage{graphicx}
\usepackage{epstopdf} 

\usepackage{amsmath}   

\begin{document}
\title{Development of Tracking Detectors with industrially produced GEM Foils}

\author{{F.~Simon, B.~Azmoun, U.~Becker, L.~Burns, D.~Crary, K.~Kearney, G.~Keeler, R.~Majka, K.~Paton, G.~Saini, N.~Smirnov, B.~Surrow, C.~Woody}
\thanks{Manuskript submitted on \today}
\thanks{F.~Simon, U.~Becker, L.~Burns and B.~Surrow are with the Laboratory for Nuclear Science, Massachusetts Institute of Technology. ({\it email: fsimon@mit.edu}).}
\thanks{B.~Azmoun and C.~Woody are with Brookhaven National Laboratory.}
\thanks{D.~Crary, K.~Kearney, G.~Keeler and G.~Saini are with Tech-Etch.}
\thanks{R.~Majka and N.~Smirnov are with the Physics Department, Yale University.}
\thanks{K.~Paton is with University of Saskatchewan.}}

\maketitle
\thispagestyle{empty}

\begin{abstract}
The planned tracking upgrade of the STAR experiment at RHIC includes a large-area GEM tracker used to determine the charge sign of electrons and positrons produced from W$^{+(-)}$ decays. For such a large-scale project commercial availability of GEM foils is necessary. We report first results obtained with a triple GEM detector using GEM foils produced by Tech-Etch Inc. of Plymouth, MA, USA. Measurements of gain uniformity, long-term stability as well as measurements of the energy resolution for X-Rays are compared to results obtained with an identical detector using GEM foils produced at CERN.  A quality assurance procedure based on optical tests using an automated high-resolution scanner has been established, allowing a study of the correlation of the observed behavior of the detector and the geometrical properties of the GEM foils. Detectors based on Tech-Etch and CERN produced foils both show good uniformity of the gain over the active area and stable gain after an initial charge-up period, making them well suited for precision tracking applications. 
\end{abstract}

\IEEEpeerreviewmaketitle

\section{Introduction}

GEM detectors are based on electron avalanche multiplication in strong electric fields created in holes etched in thin metal clad insulator foils. This concept, introduced in 1996, is referred to as the Gas Electron Multiplier (GEM) \cite{Sauli:1997qp}. Since the electron amplification occurs in the holes in the GEM foil and is separated from charge collection structures, the choice of readout geometries for detectors based on the GEM is very flexible. For tracking applications several GEM foils are cascaded to reach higher gain and high operating stability. Spatial resolutions of better than 70 $\mu$m have been demonstrated with triple GEM detectors \cite{Altunbas:2002ds}, with a material budget of significantly less than 1\% $X_0$ per tracking layer (providing a 2D space point). These features make GEM devices a natural choice for large area precision tracking, such as the planned forward tracking upgrade of the STAR detector \cite{Simon:2007NIMB}. For such a large-scale project commercial availability of GEM foils is crucial since the production capabilities of the photolithographic workshop at CERN are not sufficient. A collaboration with Tech-Etch, Inc., based on an approved SBIR proposal, has been established to provide a commercial source for GEM foils. Currently the production steps are optimized to provide GEM foils with the desired operational characteristics. Phase II of the SBIR proposal has been approved by the US Department of Energy and is currently under way. We are reporting first results obtained with triple GEM detectors using foils produced by Tech-Etch with an active area of 10 cm $\times$ 10 cm.

\begin{figure}
\centering
\includegraphics[width= 0.25\textwidth]{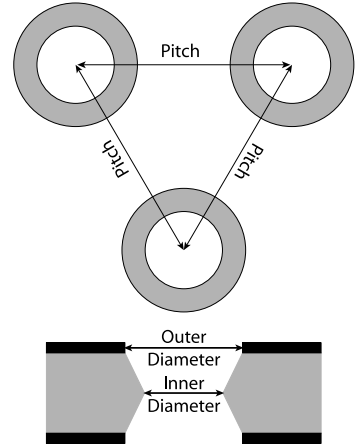}
\caption{Schematic illustration of geometrical parameters of GEM foils produced at CERN and at Tech-Etch.}
\label{fig:GEMGeometry}
\end{figure}

\section{GEM Production and Optical Tests}

\begin{figure*}
\centering
\includegraphics[width=0.9\textwidth]{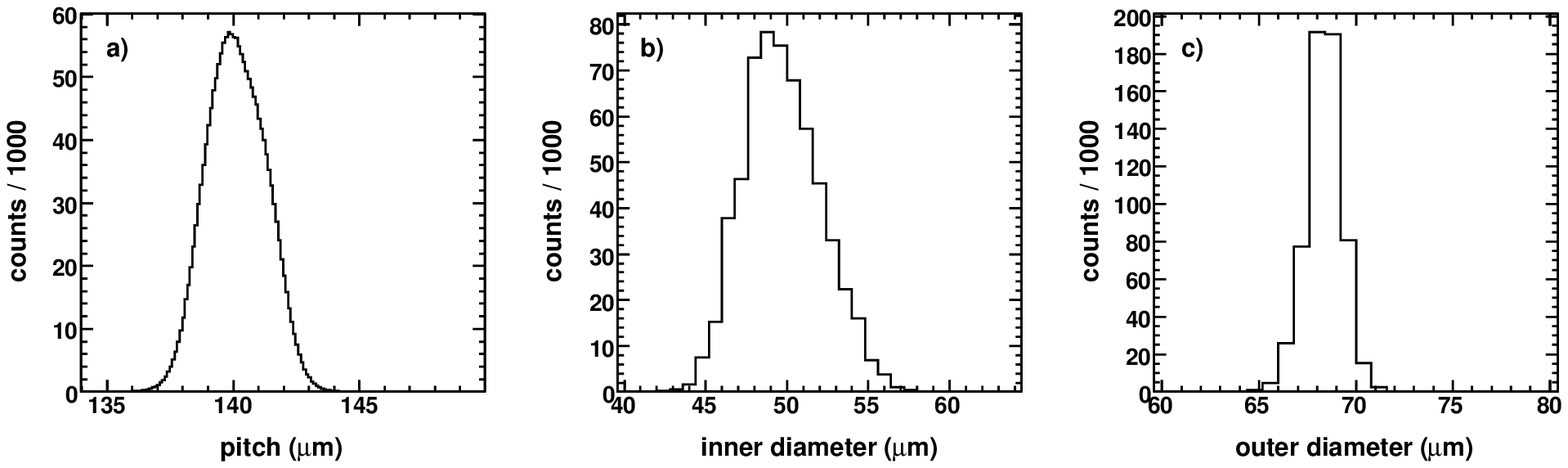}
\caption{Geometrical parameters of a typical CERN produced GEM foil determined with the optical scanner. a) shows the distribution of the hole pitch, b) the inner hole diameter (minimum diameter in the insulator layer) and c) the outer hole diameter (diameter in the copper layer).}
\label{fig:CernParameters}
\vspace{2mm}
\includegraphics[width=0.9\textwidth]{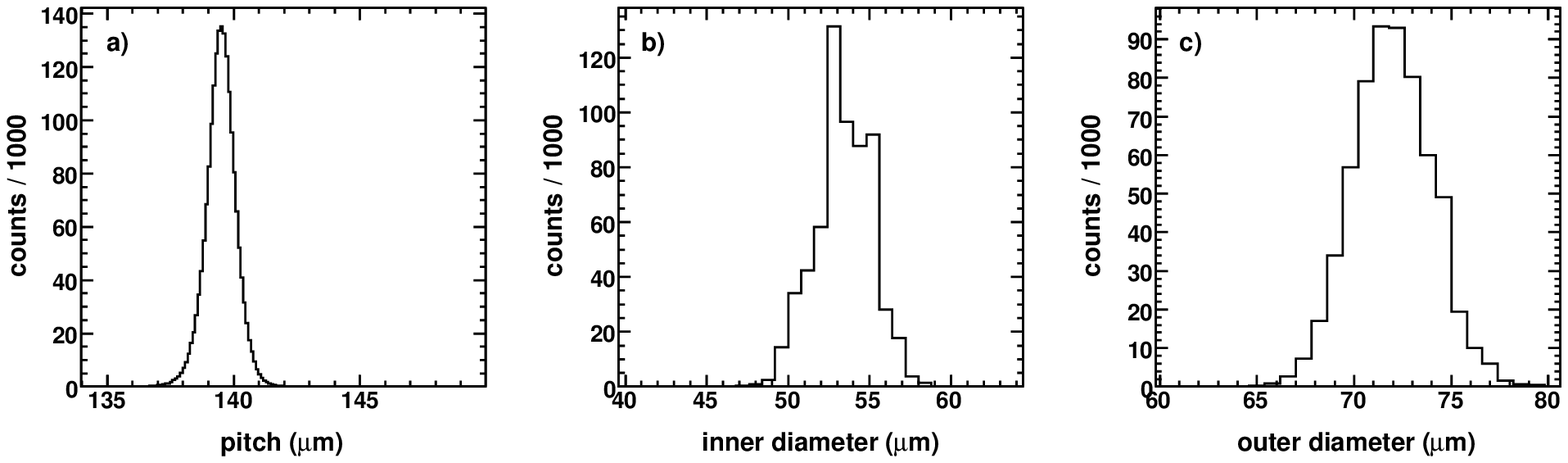}
\caption{Geometrical parameters of a typical Tech-Etch produced GEM foil determined with the optical scanner. a) shows the distribution of the hole pitch, b) the inner hole diameter (minimum diameter in the insulator layer) and c) the outer hole diameter (diameter in the copper layer).}
\label{fig:TEParameters}
\end{figure*}

Tech-Etch produces GEM foils from copper coated polyimide using photolithographic processes. After the raw material is cut to the appropriate size, a photo-resist mask is applied, imaged and developed.  The copper is then etched to form the 75 $\mu$m diameter holes, the high voltage electrodes, and the polyimide border around the GEM on the near side and far side copper layers.  The photo-mask is stripped and a second mask is applied to protect the exposed polyimide border during the etching of the polyimide GEM holes.  This second mask is removed and the GEM is complete except for a final cleaning step and electrical test to assess the GEM foilÕs quality. After these final steps, the GEM foils are distributed to the research centers for further testing and evaluation. 

An optical scanning station has been developed at MIT \cite{Becker:2006yc} to provide measurements of geometrical parameters of GEM foils. Foils from both production locations (CERN and Tech-Etch) use a standard triangular hole pattern with equidistant holes at a pitch of 140 $\mu$m and hole diameters in the metal layer of $\sim$ 70 $\mu$m. The geometry is illustrated in Figure \ref{fig:GEMGeometry}. The inner and outer hole diameters as well as the hole pitch of foils are determined with a high resolution CCD camera. A fully computer controlled setup with the camera and two dimensional stages controlled by step motors allows the examination of each individual hole. It measures the hole to hole distance (pitch) as well as the inner  and the outer diameter of each hole. Pattern recognition software is used to identify defects such as missing or blocked holes and other production defects.

GEM foils produced both by the CERN workshop and by Tech-Etch have been examined with this setup. Figure \ref{fig:CernParameters} shows the measured distributions of the hole pitch, of the inner hole diameters (minimum diameter in the insulator) and of the outer hole diameters (diameter in the copper layer) for a typical CERN produced GEM foil. The same quantities for a Tech-Etch produced foil are shown in Figure \ref{fig:TEParameters}. Foils from both producers show narrow distributions for all three quantities, indicating good geometrical uniformity of the foils.

\begin{figure}
\centering
\includegraphics[width= 0.45\textwidth]{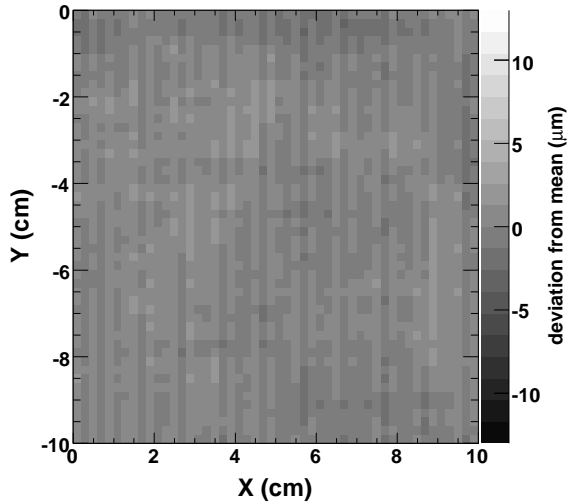}
\caption{Spatial homogeneity of the inner hole diameter of a Tech-Etch produced foil. The grey scale (indicated on the right) shows the deviation of the inner hole diameter from the mean diameter over the whole foil. The variations are on the scale of $\pm$3 $\mu$m around the mean. }
\label{fig:TEHomogeneity}
\end{figure}

The inner hole diameter is especially critical for the gain of the GEM foil and is thus of special interest. Figure \ref{fig:TEHomogeneity} shows the distribution of the inner hole diameter over the full surface area. No striking features are visible, indicating good homogeneity over the full surface area.

\section{Test Detectors}

In order to evaluate the performance of GEM foils produced by Tech-Etch in an application environment, a test detector based on the geometry used in the COMPASS experiment \cite{Altunbas:2002ds} has been developed at MIT. The detector is a triple GEM design with an active area of 10 cm $\times$ 10 cm and with a two dimensional projective strip readout. The readout structure is a laser etched printed circuit board with a two dimensional strip pattern. The strip pitch is 635 $\mu$m, the strip width for the two coordinates are chosen to achieve equal charge sharing. The foils are powered from a single high voltage source through a resistor chain with equal voltage sharing between the three foils. The resistors across the GEM foils are 1.2 M$\Omega$ each, and the resistors across the drift and transfer gaps are 2.0 M$\Omega$ each, leading to a total of 11.6 M$\Omega$. The drift gap of the detector between the cathode foil and the top GEM is $\sim$ 3 mm, the transfer gap between the other foils and between the bottom GEM and the readout board are $\sim$ 2 mm for the detector based on CERN produced foils. In the case of the Tech-Etch based detectors, different frames for the foils are used, leading to an increase of all gaps by $\sim$ 0.2 mm. For framing, the foils are stretched with a force of about two pounds and then glued to the frames. The detector is designed to allow for easy replacement of individual foils. A pre-mixed gas of Ar:CO$_2$ (70:30) is used for all measurements, and the detectors are operated at ambient pressure. For measurements with X-Rays the test detectors are read out via a charge sensitive preamplifier (Cremat CR-110 \cite{cremat}) and a shaping amplifier (Ortec 571, shaping time 0.5 $\mu$s).  For these studies several readout strips are combined into one channel, forming a single active readout area roughly 1.2 cm wide and 10 cm long. The data acquisition is based on a PC controlled CAMAC system, using a standard CAMAC ADC (LeCroy 2249W). A strip-by-strip readout based on the APV25-S1 front-end chip \cite{French:2001xb} has also been developed and was successfully used in tests with particle beams at the MTest test beam facility at Fermilab. The results of these tests will be discussed in a later publication. 

\begin{figure}
\centering
\includegraphics[width=0.48\textwidth]{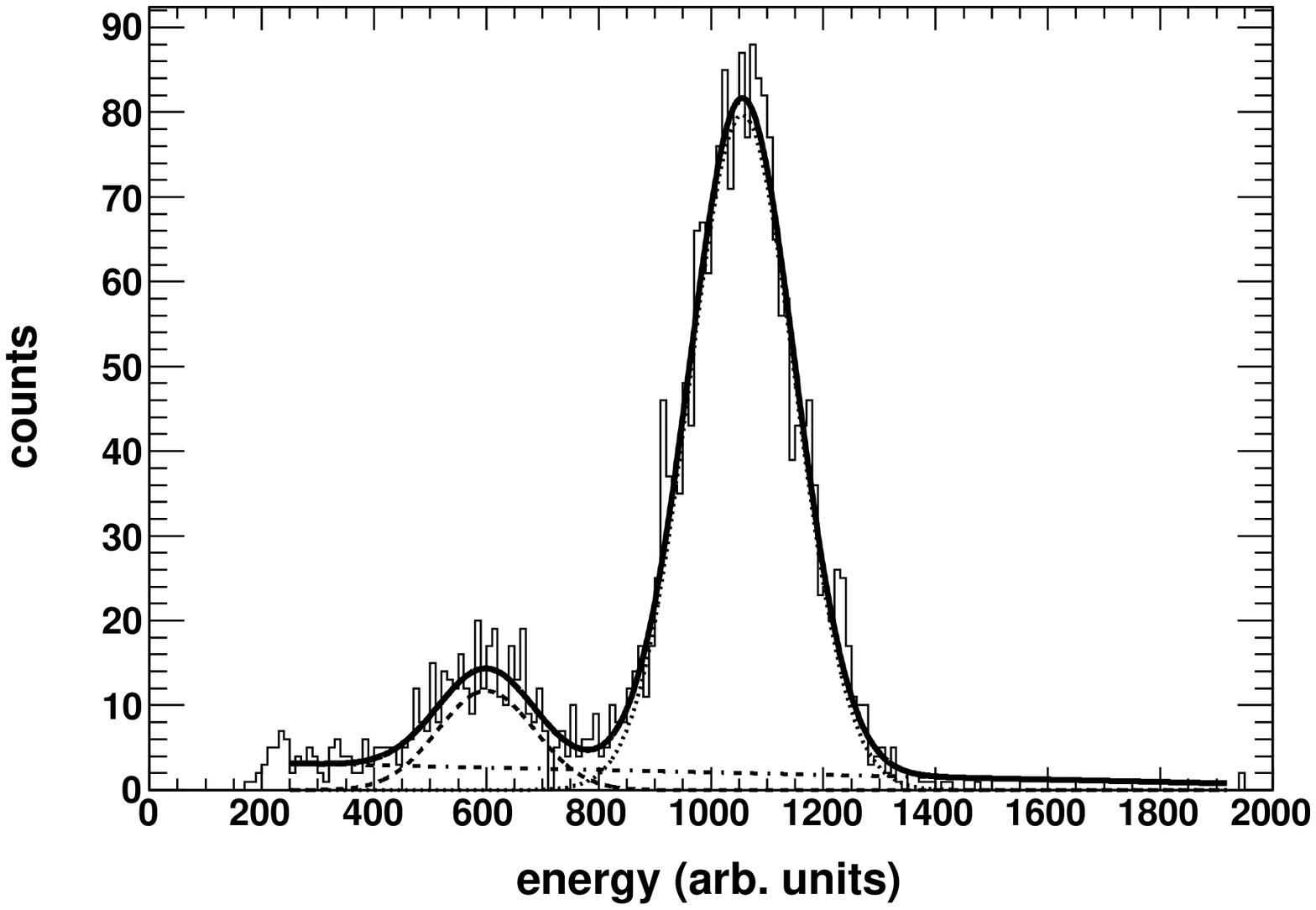}
\caption{$^{55}$Fe spectrum (main line at 5.9 keV) taken with a triple GEM test detector using CERN GEM foils, operated at a gain of $\sim$ 25\,000. The spectrum is fitted with the sum of two Gaussians and a linear background. The energy resolution (FWHM of the photo peak divided by the mean) is $\sim$18\%.}
\label{fig:CernSpec}
\includegraphics[width=0.48\textwidth]{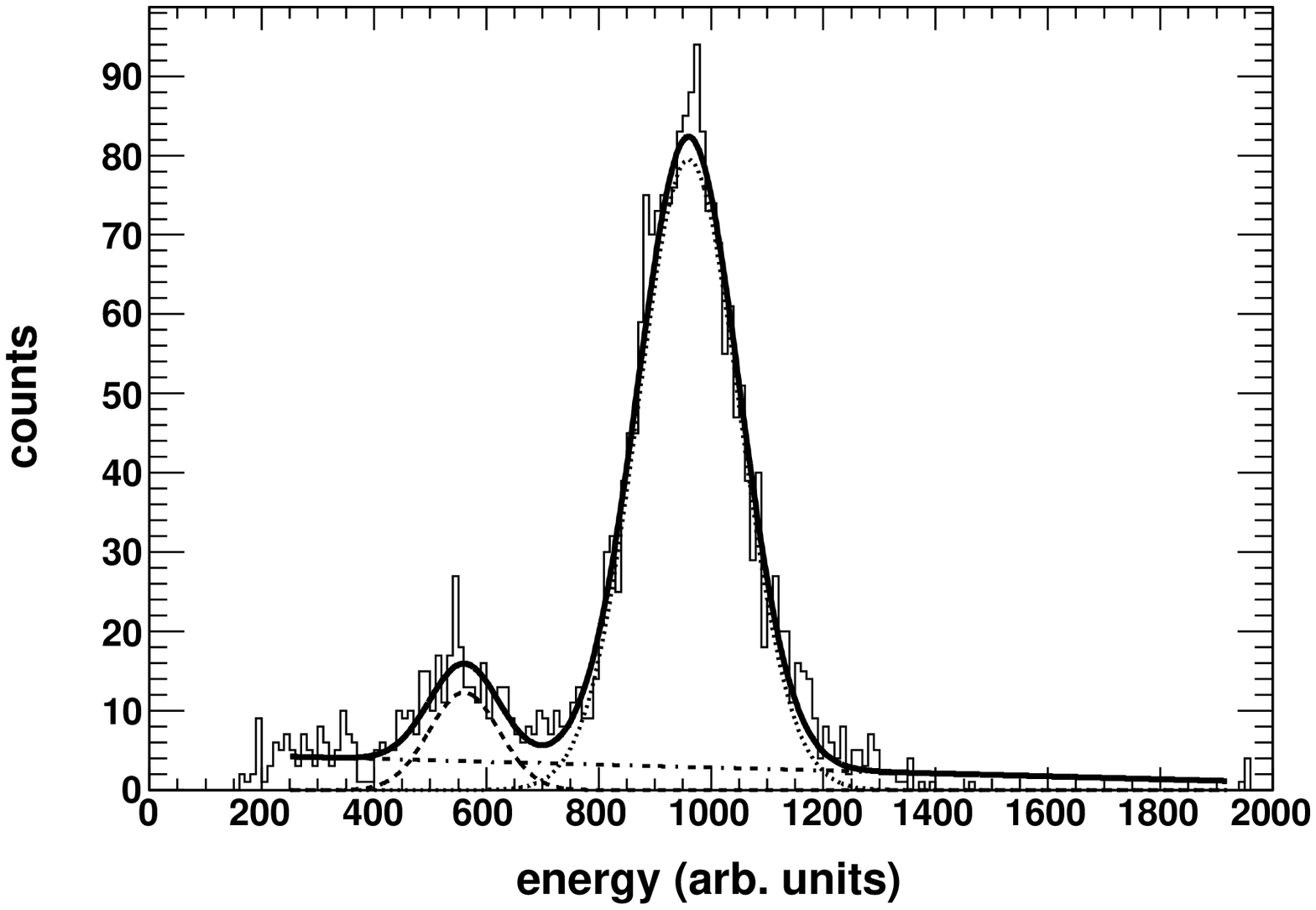}
\caption{$^{55}$Fe spectrum (main line at 5.9 keV) taken with a triple GEM test detector using Tech-Etch  GEM foils, operated at a gain of $\sim$ 25\,000. The energy resolution is $\sim$19\%.}
\label{fig:TESpec}
\end{figure}

First studies comparing a detector constructed with CERN made GEM foils with one based on Tech-Etch foils have been made using a low-intensity $^{55}$Fe source (mainly 5.9 keV photons) with a rate of $\sim$0.5 Hz/mm$^2$. Figure \ref{fig:CernSpec} shows a typical spectrum recorded with the CERN foil based detector, while Figure \ref{fig:TESpec} shows a spectrum recorded with the detector using Tech-Etch GEM foils. The measurements where taken after the high voltage had been turned on for a longer period (several hours) and after the detectors had been exposed for a period of several minutes to a higher radiation intensity of several kHz/mm$^2$ over the full active area. This leads to a full charging up of the detectors, as discussed in Section \ref{sec:Charging}. For the present measurements the detectors were operated at voltages of around 400 V across each GEM foil, with transfer and induction fields of $\sim$3.3 kV/cm and a drift field of $\sim$2.2 kV/cm. The effective gain of the detectors for these measurements was around $2.5 \times 10^4$, as discussed in more detail in the next section. The voltages were adjusted for each detector individually to use the full dynamic range of the readout system. The Tech-Etch based detector needed about 200 V higher overall voltage, corresponding to 20.7 V more across each GEM, than the CERN detector to achieve a similar gain. This is assumed to be due to slightly larger hole diameters of the Tech-Etch foils. In both the CERN and the Tech-Etch detector a clean separation of the main photo peak and the Ar escape peak is achieved. The energy resolution, defined by the ratio of the photo peak FWHM and the mean of the peak, is on the order of 20\% for both detectors.

\begin{figure}
\centering
\includegraphics[width=0.45\textwidth]{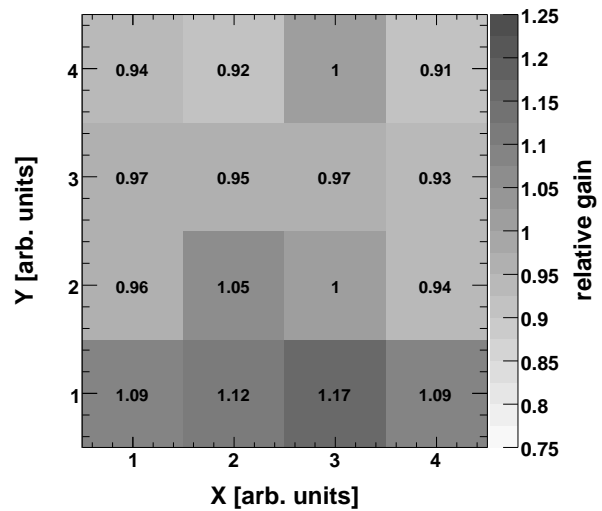}
\caption{Map of the relative gain as a function of spatial location for a triple GEM detector using Tech-Etch produced GEM foils. The relative gain, normalized to the mean, is shown by the color scale and indicated by the numbers in each segment.}
\label{fig:TEBNLGainDistribution}
\end{figure}

\begin{figure}
\centering
\includegraphics[width=0.45\textwidth]{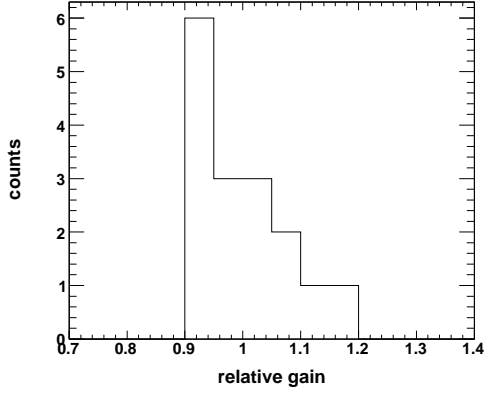}
\caption{Distribution of the relative gains over the detector area, as shown in the gain map in Figure \ref{fig:TEBNLGainDistribution}. The variance of the distribution is 0.078, indicating good uniformity of the detector gain over the active area.}
\label{fig:TEBNLGainHist}
\end{figure}

By measuring the $^{55}$Fe pulse height in 16 different places, effectively dividing the active area in a $4\times 4$ grid, a map of the relative gain as a function of spatial location is obtained. Figure \ref{fig:TEBNLGainDistribution} shows the gain distribution over the 10 cm $\times$ 10 cm active are of a triple GEM test detector using Tech-Etch produced GEM foils. Figure \ref{fig:TEBNLGainHist} shows a histogram of the 16 relative gains measured over the active area. The small RMS of 0.078 indicates a good uniformity of the detector. Only two out of the 16 measured gains are more than 10\% off of the mean value. Similar observations were also made with a test detector using CERN produced GEM foils. These results are in line with the observations made with the COMPASS triple GEM detectors in similar measurements \cite{Altunbas:2002ds}.

\section{Gain Evolution with Voltage, Temperature, and Pressure}

\begin{figure}
\centering
\includegraphics[width=0.45\textwidth]{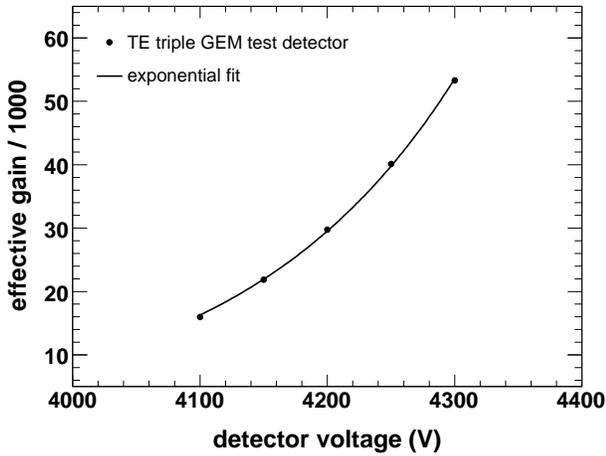}
\caption{Effective gain of a Tech-Etch foil based triple GEM detector as a function of overall detector voltage. The gain shows the expected exponential behavior. 4200 V correspond to 433 V per GEM. The uncertainty of the gain measurement is estimated to be 20\%.}
\label{fig:GainvsVoltage}
\end{figure}

The gain of a gaseous detector is a function of both the applied voltage and of the density of the detector gas, which in turn depends on the temperature and the pressure. In this section, a detailed study of the gain of a triple GEM detector using Tech-Etch produced GEM foils is presented. Figure \ref{fig:GainvsVoltage} shows the dependence of the overall effective detector gain as a function of applied voltage. The voltage across the individual GEM foils is related to the overall detector voltage by
\begin{equation}
V_{GEM} = 0.103 \times V_{Detector}.
\end{equation}
The gain is determined from measurements of the photo peak position of $^{55}$Fe spectra at different voltages. The charge sharing between the two readout coordinates, where only one is read out, is taken into account in the calculation of the effective gain. The precision of the gain determination, including the calibration of the used amplifier setup with capacitive charge injection, is estimated to be 20\%. Due to limited shielding and the large capacitance of the group of readout strips the noise on the signal was very high, requiring a high-gain operation of the detectors. The gain shows the expected exponential dependence on the applied voltage. The high effective gain that is achieved with Tech-Etch triple-GEM detectors demonstrates good performance and high robustness of the GEM foils.

\begin{figure}
\centering
\includegraphics[width=0.485\textwidth]{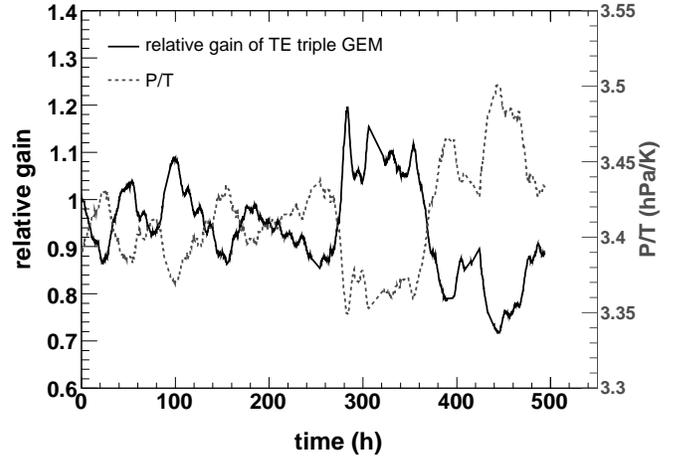}
\caption{Evolution of relative gain and P/T as a function of time for a long-term (21 days) test of the a Tech-Etch triple GEM. The P/T variations are mainly due to changes in the atmospheric pressure. }
\label{fig:GainTPvsTime}
\end{figure}

A Tech-Etch detector was operated for an extensive period to study the evolution of the gain with ambient temperature and pressure. The detector was constantly irradiated with a low intensity $^{55}$Fe source. 16\,000 events are accumulated per data point, corresponding to one point every $\sim$ 400 s. The gain for each point is determined from the photo peak position of  the spectrum. Figure \ref{fig:GainTPvsTime} shows the variation of the effective detector gain and the ratio of pressure and absolute temperature $P/T$ as a function of time over a period of three weeks, starting after the initial charging and gain stabilization. The anti-correlation of the gain with $P/T$ is clearly apparent in the figure. The variations in $P/T$ are mostly due to variations of the atmospheric pressure, since the temperature in the laboratory was stable within about 2 K. 

\begin{figure}
\centering
\includegraphics[width=0.485\textwidth]{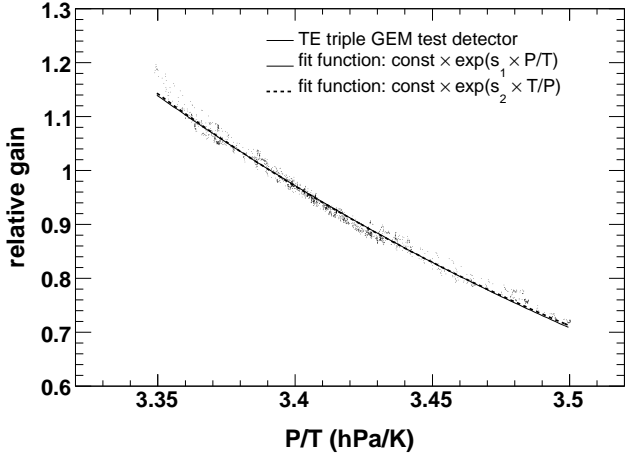}
\caption{Relative detector gain as a function of $P/T$, obtained from gain measurements over a period of three weeks exploiting variations in atmospheric pressure. An exponential function fitted to the distribution yields a slope of -3.17 K/hPa.}
\label{fig:GainvsTP}
\end{figure}

The gain of gaseous detectors is described by the first Townsend coefficient, which is the inverse of the mean free path for ionization, the average distance an electron has to travel before participating in an ionizing collision. This in turn is a function of the gas density, which depends on temperature and pressure. The dependence of the Townsend coefficient on the gas density is non-trivial and is discussed in greater detail for example in \cite{Aoyama:1985GasGain} and references therein. Here only a very limited range in temperature and pressure around standard conditions is relevant. We are thus investigating two commonly used simplified parameterizations,  one where the Townsend coefficient is assumed to be proportional to $P/T$, and one where it is assumed to be proportional to $T/P$ in the range of parameters relevant to the measurement. This results in the following two functional dependencies for the detector gain that are considered:
\begin{equation}
G = const \times e^{s_1 P/T} \quad \mbox{and} \quad
G  = const \times e^{s_2 T/P}, 
\end{equation}
where $s_1$ and $s_2$ are the slopes of the exponential functions.

Figure \ref{fig:GainvsTP} shows the relative gain as a function of $P/T$ for the time period shown in Figure \ref{fig:GainTPvsTime}. The distribution is well described by an exponential function in $P/T$ with a slope of \mbox{$-3.17$ K/hPa}, as well as by an exponential function in $T/P$ with a slope of 36.9 hPa/K. The observed slope parameter for the exponential function in $T/P$ is consistent with observations made with the COMPASS triple GEM detectors \cite{Altunbas:2002ds}. Based on these values a correction for environmental effects on the observed gain is possible. An automatic adjustment of the detector voltage to stabilize the gain might be achievable, depending on the size of charge up effects, as discussed in Section \ref{sec:Charging}. For the planned application in tracking detectors for the STAR experiment such a gain stabilization is very likely not necessary.

\section{Gain Increase through Charge Deposition}
\label{sec:Charging}

\begin{figure}
\centering
\includegraphics[width=0.48\textwidth]{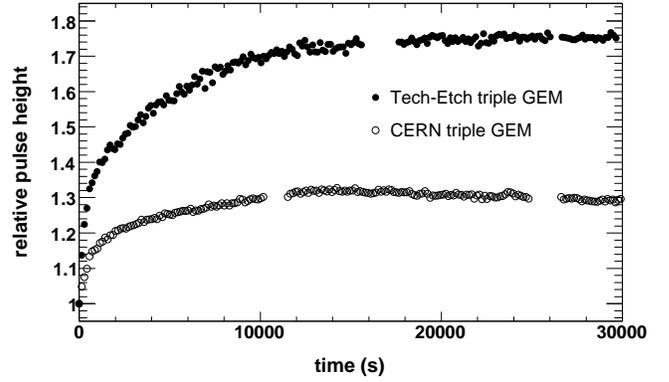}
\caption{Evolution of the relative gain for the Tech-Etch and CERN foil based detector as a function of time during irradiation with a low intensity $^{55}$Fe source ($\sim$0.5 Hz/mm$^2$). The pulse height of the first measurement with each detector is normalized to 1. No corrections for temperature and pressure are applied. While environmental parameters were stable during the Tech-Etch measurement, the ambient pressure changed during the measurement with the CERN detector, leading to small variations of the gain in the plateau region.}
\label{fig:Gain}
\end{figure}

Charge deposition on the insulator within the GEM holes and polarization of the polyimide leads to modifications of the electric field and thus to changes in the detector gain over time. This is a well-known phenomenon in GEM based detectors \cite{Bouclier:1997pg}. Figure \ref{fig:Gain} shows the evolution of the detector gain over an extended period of time after turn on for both a CERN and a Tech-Etch triple GEM detector under constant irradiation with a low intensity $^{55}$Fe source ($\sim$0.5 Hz/mm$^2$). Both detectors show a gain increase after turn-on, but reach a stable gain plateau within approximately 3 hours. The charging up can be accelerated by exposing the detector to higher intensity radiation. The significantly larger gain increase of the Tech-Etch detector is attributed to larger insulator areas exposed in the holes, as compared to the CERN produced foils. This effect is discussed in more detail below. The slightly different geometry of the Tech-Etch detector leads to a 10\% change of the ratio of the fields inside the GEM foils to the transfer and induction fields, which can also increase the initial charging effects, as reported in \cite{Bachmann:1999xc, Bondar:2005uc}.

\begin{figure}
\centering
\includegraphics[width=0.48\textwidth]{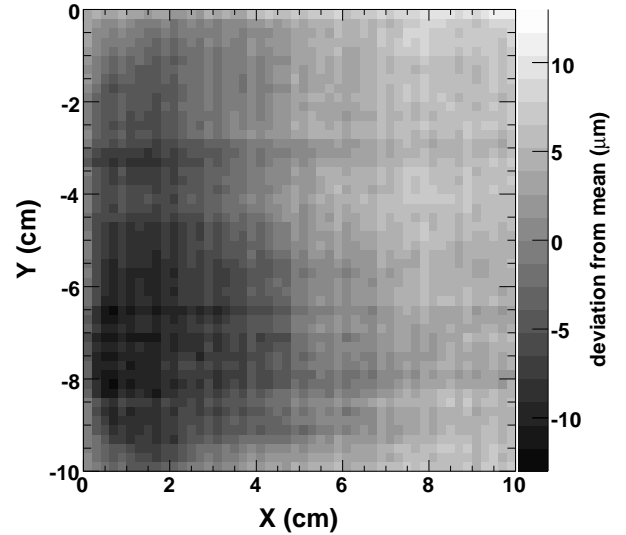}
\caption{Spatial homogeneity of the inner hole diameter of a Tech-Etch produced foil that shows large variations over the active area. The grey scale (indicated on the right) shows the deviation of the inner hole diameter from the mean diameter over the whole foil. The inner diameter of holes in the lower left corner of the foil is $\sim$20 $\mu$m smaller than on the right side of the foil.}
\label{fig:BadInnerHole}
\end{figure}

\begin{figure}
\centering
\includegraphics[width=0.48\textwidth]{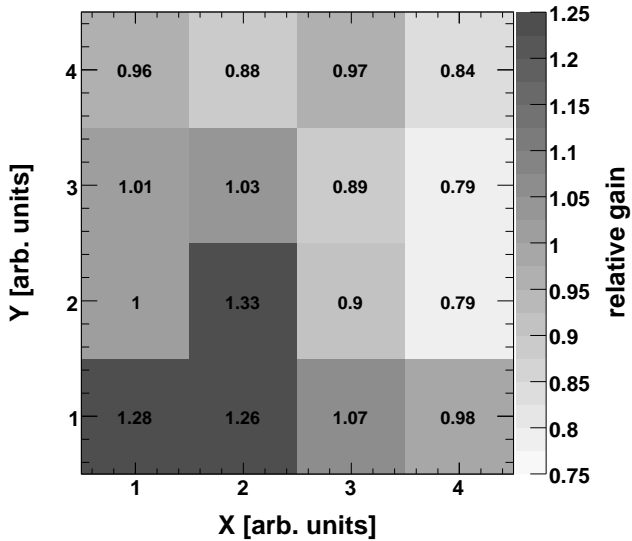}
\caption{Map of the relative gain as a function of spatial location for a triple GEM detector using the bad Tech-Etch foil shown in Figure \ref{fig:BadInnerHole} as top foil, replacing the homogeneous foil shown in Figure \ref{fig:TEHomogeneity}. The relative gain, normalized to the mean is shown by the color scale and indicated by the numbers in each segment. The distribution of the relative gains has an RMS of 0.16.}
\label{fig:BadGainMap}
\end{figure}

It is known that the hole geometry affects the charging up behavior of GEM foils. More cylindrical geometries (e.g. inner hole radius close to outer hole radius) show less charging up than conical or extreme double conical geometries with a large insulator surface exposed in the holes \cite{Bouclier:1997pg,Benlloch:1998wa}. 

This was investigated by using a GEM foil with a large non-uniformity in the inner hole diameter, but with uniform outer hole diameters. The optical scan of such a foil from Tech-Etch production is shown in Figure \ref{fig:BadInnerHole}. The large non-uniformities were caused by problems in the insulator etching phase, thus the outer hole diameters were not affected. 

Figure \ref{fig:BadGainMap} shows the gain map of a triple GEM detector where the original top foil in the three foil stack, shown in Figure \ref{fig:TEHomogeneity}, has been replaced by this non-uniform foil. This measurement was done with the detector fully charged up. Comparison to the gain map with homogeneous foils shown in Figure \ref{fig:TEBNLGainDistribution} shows the dramatic changes introduced by the problematic top foil. Since the measurements are done with triple GEM detectors, effects from the two lower GEM foils also contribute. These are the same in Figures \ref{fig:TEBNLGainDistribution} and \ref{fig:BadGainMap}. In the areas with smaller inner holes the gain is significantly increased over the average. Before charging up, the situation is different. The gain in the area of small holes, measured at $x$ $\sim$ 2 cm and $y$ $\sim$ $-8$ cm, as shown in Figure \ref{fig:BadInnerHole}, is initially lower than the gain in the area of larger holes at $x$ $\sim$ 7 cm and $y$ $\sim$ $-4$ cm. Without irradiation, the gain increases by about 12\% in the area of small holes and less than 5\% in the areas of the larger holes over the period of one hour. After intense irradiation, more than a doubling of the gain in the small hole area was observed, while the gain in the areas with larger inner hole diameters increased by about 50\%. These results, together with previous observations with different hole geometries \cite{Benlloch:1998wa} demonstrate that a reduction of the difference between inner and outer hole diameter and thus the exposed insulator surface in the holes is desirable to reduce the charging up effect. Almost cylindrical holes however tend to lead to significantly increased discharge rates due to sharp metal edges exposed in the high-field regions inside the GEM holes. These edges get formed due to over-etching of the insulator material underneath the metal layers. Thus an optimum production setup has to be found to obtain foils with a large ratio of inner to outer hole diameter while excluding the possibility of over-etching. 

\section{Conclusion}

Comparative measurements of geometrical parameters of CERN and Tech-Etch produced GEM foils and first results obtained with a triple GEM detector using foils produced by Tech-Etch, Inc. have been reported. The foils from both manufacturers had an active area of 10 cm  $\times$ 10 cm. The geometric parameters of Tech-Etch and CERN produced foils are found to be very similar. Foils from both manufacturers show good uniformity of the parameters over the full area of the foils. The energy resolution of triple GEM test detectors using Tech-Etch foils for $^{55}$Fe X-Rays is comparable to that measured with a reference detector using GEM foils manufactured at CERN.  Detectors based on Tech-Etch and CERN produced foils show similar gain uniformity over the active area. The achieved uniformity is sufficient for the planned applications in the STAR tracking upgrade. 
High gains in excess of $5 \times 10^4$ where achieved with detectors using Tech-Etch foils in stable long-term operation in the presence of low-intensity X-ray irradiation. The variation of the gain with temperature and pressure can be described by a simple exponential function in a narrow range around standard conditions. The observed variations are consistent with observations made with the COMPASS triple GEM detectors. Tech-Etch based detectors show a larger initial gain increase due to charging up than their CERN based counterparts. Both detector types reach a stable gain plateau after charging up, thus this behavior does not affect the suitability of the detectors for tracking applications. The charging up is at least in part due to the hole geometry, with holes with a smaller ratio of inner to outer hole diameter leading to more charging up. In the ongoing phase II of the SBIR program an optimized production process for GEM foils is being developed to further improve foil uniformity and performance.

\section*{Acknowledgments}

The authors thank F.~Sauli from INFN Trieste and CERN for helpful suggestions and discussions. The development of GEM foil production at Tech-Etch is supported by US-DOE SBIR grant DE-FG02-05ER84169.

\bibliographystyle{IEEEtran.bst}
\bibliography{GEM}

\end{document}